\documentclass[english,aps,prd,reprint,amsfonts,amsmath,amssymb,superscriptaddress]{revtex4-1}
\usepackage[T1]{fontenc}
\usepackage[latin9]{inputenc}
\usepackage{verbatim}
\usepackage{amsmath}
\usepackage{stackrel}
\usepackage{graphicx}

\makeatletter
\usepackage{hyperref}

\makeatother

\usepackage{babel}
\begin{document}

\title{First-order post-Newtonian analysis of the relativistic tidal effects
for satellite gradiometry and the Mashhoon-Theiss anomaly}

\author{Peng Xu}

\email{xupeng@amss.ac.cn}

\affiliation{Institute of Applied Mathematics, Academy of Mathematics and Systems
Science, Chinese Academy of Sciences, 55 Zhongguancun Donglu, Beijing
100190, China}

\affiliation{Morningside Center of Mathematics, Chinese Academy of Sciences, 55
Zhongguancun Donglu, Beijing 100190, China}

\author{Ho Jung Paik}

\email{hpaik@umd.edu}

\affiliation{Department of Physics, University of Maryland, College Park, MD 20742, USA}

\begin{abstract}
With continuous advances in technology, future satellite gradiometry missions
will be capable of performing precision relativistic experiments and
imposing constraints on modern gravity theories. To this end, the full
first-order post-Newtonian tidal tensor under inertially guided and
Earth-pointing local frames along post-Newtonian orbits is worked out.
The physical picture behind the ``Mashhoon-Theiss anomaly'' is explained
at the post-Newtonian level. The relativistic precession of the
local frame with respect to the sidereal frame will produce modulations
of Newtonian tidal forces along certain bases, which gives rise to
two different kinds of secular tidal tensors. The measurements of
the secular tidal force from the frame-dragging effect is also discussed.
\end{abstract}
\maketitle

\section{Introduction\label{sec:Introduction}}

Gradiometers, measuring the gradients of gravitational force, are
capable of reading out directly the local functionals of spacetime curvature.
Today, gradiometry has found wide applications, especially in geodesy.
The GOCE satellite, as an example, carried a 3-axis electrostatic
gradiometer to map the geopotential of Earth, whose sensitivity
already reached about 10 mE Hz$^{-1/2}$ in the band of $5 \sim 100$
mHz, where 1 E $\equiv 10^{-9}$ s$^{-2}$
\cite{Rummel2011}. With the development of more advanced
superconducting \cite{Griggs2015} and atomic gradiometers \cite{Tino2013,Carraz2014},
satellite gradiometry could be employed for precision relativistic
experiments to impose constraints on
modern theories of gravitation. As a potential outstanding test of
Einstein's general theory of relativity (GR), the studies of the detection
of gravitomagnetic (GM) effects with orbiting gradiometers
started in the 1980s \cite{Paik1988,Mashhoon1989}.

Within the weak field and slow motion limit, there exists rich correspondence
between GR and Maxwell's electrodynamics \cite{Thorne1988,Maartens1998,Mashhoon2003}.
In fact, due to the Lorentz symmetry pertaining to the flat background
spacetime, the analogous GM field arising from mass currents
shows up in general metric theories of gravity \cite{Dicke1964,Will2014}.
The detection of such GM effects can be related to fundamental issues,
such as quantitative measurements of how local inertial frames are
altered by nearby massive sources (Mach's principle) \cite{Sciama1953,Ciufolini1995}.

In 2011, NASA's GP-B team announced detection of the Earth GM field with 20\%
accuracy \cite{Everitt2011} by measuring the precession of
freely falling gyros \cite{Schiff1960} with respect to the
direction of a guide star, HR 8703.
Another experiment is to trace the relative precession of the ascending nodes of the
twin satellites LAGEOS I and II \cite{Lense:1918zz,Ciufolini1986}.
Combining with the gravity field model provided by the CHAMP \cite{CHAMP}
and GRACE missions \cite{GRACE},
Ciufolini and collaborators also reported confirmation of the GM
effect with the precision of 10\% \cite{Ciufolini2004,Ciufolini2007}.
Different from these methods, Braginskii and Polnarev in 1980s derived
the coupling  between orbiting oscillators and the GM field of
a rotating mass \cite{Braginskii1980}. It is soon noticed by
Mashhoon, Paik and collaborators that such effect could be detected
by an orbiting 3-axis gradiometer, which could fit naturally in a
satellite gravity mission \cite{Mashhoon1982,Paik1988,Paik1989,Mashhoon1989}.
Having laser interferometry as an alternative readout method for gradiometry,
like in the LISA PathFinder mission \cite{LPF}, a new measurement scheme to detect
planetary GM effects is also under investigation \cite{Xu2016}.

Along this line, Mashhoon and Theiss claimed to have found a new relativistic
nutational oscillation in the axes of gyroscopes orbiting a rotating
source due to a small divisor phenomenon involving the geodetic precession
frequency \cite{Mashhoon1984,Mashhoon1985}. Such effect, known as
the ``Mashhoon-Theiss anomaly'', could also manifest itself in the motion
of the Earth-Moon system \cite{Mashhoon1986}, and more importantly
in a ``resonance-like'' way in certain tidal tensor components
\cite{Mashhoon1982,Theiss1985}.
Before long, it was shown \cite{Gill1989}
that such ``anomaly'' cannot be a new relativistic (resonance) effect
related to rotating sources but appears already at the post-Newtonian
(PN) level due to an interplay between the geodetic \cite{Schiff1960}
and Lense-Thirring (LT) precession \cite{Lense:1918zz}. Part of the
secular tidal forces were also computed with this approach \cite{Gill1989,Blockley1990}.
Today confusions or even disagreements about this issue still remain
in the literature. Therefore, it is necessary to work out the \textit{complete
PN tidal tensor} and analyze the possible secular terms appearing \textit{at
the 1PN level} for future satellite gradiometry missions.

This paper is continuation to those former works,
in particular, completion to the classical work by Mashhoon, Paik and Will
\cite{Mashhoon1989}, with the PN perturbations of the orbit and the PN precession of
the local frame included. With the settings explained in Sec. \ref{sec:Settings},
the complete 1PN tidal tensors, especially the secular tensors, are
derived in Sec. \ref{sec:The-Fermi} under the inertially guided frame
of the spacecraft (S/C). Three kinds of relativistic precession,
that is, the LT precession of the orbit, the geodetic, and the
Schiff (frame-dragging) precession of freely falling frames, will alter
gradually the relative orientation of the local frame with respect
to the sidereal frame, thereby producing secular changes in certain
tidal tensor components through the modulations of Newtonian tidal
forces. These relativistic secular terms can be divided into two parts,
one from the geodetic precession and the other from the combined action
of the LT and Schiff precession. The results for Earth-pointing
orientation are also worked out in Sec. \ref{sec:Earth-pointing}.
Finally, in Sec. \ref{sec:Conclusions}, the measurements of the 1PN
secular tidal forces with an on-board 3-axis gradiometer is briefly discussed.

\section{Coordinate system and post-Newtonian orbit\label{sec:Settings}}

The PN coordinate system $\{t,\vec{x}\}$ outside a rotating source
$M$ is chosen as follows. The mass center of the source is set at
the origin. The coordinate basis $\frac{\partial}{\partial x^{3}}$
is set to be parallel with the direction of the source's angular momentum
$\vec{J}$, $\frac{\partial}{\partial x^{1}}$ is pointing to a reference
star $\Upsilon$, and $\frac{\partial}{\partial x^{2}}$ is determined
by the right-hand rule $\frac{\partial}{\partial x^{1}}\times\frac{\partial}{\partial x^{2}}=\frac{\partial}{\partial x^{3}}$.
Such a coordinate system is tied to remote stars, and $t$ is the
time measured by the observer at asymptotically flat region.
Geometrized units,  $c=G=1$, are adopted hereafter. Here we model
the source as an ideal and uniform rotating spherical body.

In this coordinate system, the PN metric outside the source has
the form \cite{Weinberg1972,Straumann1984}:
\begin{equation}
g_{\mu\nu}=\left(\begin{array}{cccc}
-1+2U-\frac{2M^{2}}{r^{2}} & \frac{2x^{2}J}{r^{3}} & -\frac{2x^{1}J}{r^{3}} & 0\\
\\
\frac{2x^{2}J}{r^{3}} & 1+\frac{2M}{r} & 0 & 0\\
\\
-\frac{2x^{1}J}{r^{3}} & 0 & 1+\frac{2M}{r} & 0\\
\\
0 & 0 & 0 & 1+\frac{2M}{r}
\end{array}\right),\label{eq:metri}
\end{equation}
where $U=\frac{M}{r}$ is the Newtonian potential, $r=\sqrt{\stackrel[n=1]{3}{\sum}(x^{n})^{2}}$,
and $M$ and $J$ are the total mass and angular momentum of the source.
Deviations from uniform sphere of the source will add multiple
corrections to the Newtonian potential in the above metric, whose
effects can be found in \cite{Li2014}, and will be ignored in this paper.

For a test mass (or S/C) that is orbiting around Earth with velocity $v$,
one has basic order relations:
\begin{subequations}\label{eq:orders}
\begin{eqnarray}
v^{2} & \sim & \frac{M}{r}\sim\mathcal{O}(\epsilon^{2})\ll1,\label{eq:MJorder1}\\
v^{4} & \sim & \frac{M^{2}}{r^{2}}\sim\frac{Jv}{r^{2}}\sim\mathcal{O}(\epsilon^{4}),\label{eq:MJorder2}
\end{eqnarray}
\end{subequations}
and $\epsilon\leq10^{-5}.$ Up to the 1PN level, the equation of motion
of a freely falling test mass $m$ reads
\begin{equation}
m\frac{d^{2}\vec{x}}{dt^{2}}=\vec{F}_{N}+\vec{F}_{GE}+\vec{F}_{GM},\label{eq:geodesic}
\end{equation}
where $\vec{F}_{N}$ is the Newtonian force, and the 1PN gravitoelectric
(GE) force and GM Lorentz force have the form \cite{Petit2010}:
\begin{eqnarray}
\vec{F}_{GE} & = & \frac{mM}{r^{3}}[(\frac{4M}{r}-v^{2})\vec{x}+4(\vec{x}\cdot\vec{v})\vec{v}],\label{eq:FGE}\\
\vec{F}_{GM} & = & 2m\vec{v}\times[\frac{\vec{J}}{r^{3}}-\frac{3(\vec{J}\cdot\vec{x})\vec{x}}{r^{5}}].\label{eq:FGM}
\end{eqnarray}
For simplicity, we assume that our S/C follows an almost circular orbit.
Then, from the above equation of motion, we find
{\small
\begin{subequations}\label{eq:orbit}
\begin{eqnarray}
x^{1} & = & a\cos\Psi\cos\left(\frac{2J\tau}{a^{3}}\right)-a\cos i\sin\Psi\sin\left(\frac{2J\tau}{a^{3}}\right),\label{eq:x1}\\
x^{2} & = & a\cos i\sin\Psi\cos\left(\frac{2J\tau}{a^{3}}\right)+a\cos\Psi\sin\left(\frac{2J\tau}{a^{3}}\right),\label{eq:x2}\\
x^{3} & = & a\sin i\sin\Psi,\label{eq:x3}
\end{eqnarray}
\end{subequations}}where the initial longitude of the ascending node $\Omega$ is
set to
be zero, $i$ is the inclination angle, $a$ is the radius and $\Psi=\omega\tau$
with $\tau$ the proper time measured by the on-board clock and $\omega$
the orbital frequency. The orbital plane of the S/C will slowly precess
due to the frame-dragging effect with a period of $\sim 10^{7}$ yr for
medium altitude Earth orbits. The effect of small eccentricity and other orbital
perturbations will be left to future studies.

For two adjacent freely orbiting test masses, we introduce a vector
$Z^{\mu}$ pointing from the reference mass to the second. For the
case of $d=|Z|\sim10^{-1}$ m, which is much shorter than the orbital
semi-major $a\sim10^{7}$ m, the relative motion between the test
masses can be obtained by integrating the geodesic deviation equation:
\begin{equation}
\tau^{\rho}\nabla_{\rho}\tau^{\lambda}\nabla_{\lambda}Z^{\mu}+R_{\rho\nu\lambda}^{\ \ \ \ \mu}\tau^{\rho}\tau^{\lambda}Z^{\nu}=0.\label{eq:deviation}
\end{equation}
Here $\tau^{\mu}$ denotes the 4-velocity of the reference mass and
$R_{\rho\nu\lambda}^{\ \ \ \ \mu}$ is the Riemann curvature tensor.
We use $i,j,k,...=1,2,3$ to index the spatial tensor components and
$\mu,\nu,\lambda,...=0,1,2,3$ the spacetime tensor components. We
introduce the local tetrad carried by the reference mass $e_{(a)}^{\ \ \ \mu},\ a=0,1,2,3$,
with $e_{(0)}^{\ \ \ \mu}=\tau^{\mu}$, which determines the co-moving
local frame of the gradiometer. $e_{(a)}^{\ \ \ \mu}$ can be viewed
as the transformation matrix from the local frame to the global PN
system, and the inverse matrix is denoted as $e_{\ \ \ \mu}^{(a)}$.
The geodesic deviation equation, Eq. (\ref{eq:deviation}), can be expanded
in such a local frame as
\begin{eqnarray}
 &  & \frac{d^{2}}{d\tau^{2}}Z^{(a)}\nonumber \\
 & = & -2\gamma_{\ \ (b)(0)}^{(a)}\frac{d}{d\tau}Z^{(b)}-(\frac{d}{d\tau}\gamma_{\ \ (b)(0)}^{(a)}+\gamma_{\ \ (b)(0)}^{(c)}\gamma_{\ \ (c)(0)}^{(a)})Z^{(b)}\nonumber \\
 &  & -K_{(b)}^{\ (a)}Z^{(b)}.\label{eq:localdev}
\end{eqnarray}
where $\gamma_{\ (b)(c)}^{(a)}=e^{(a)\nu}e_{(b)\nu;\mu}e_{(c)}^{\ \ \ \mu}$
are the Ricci rotation coefficients \cite{Wald1984} and $;$ denotes the covariant
derivative associate to the metric, Eq. (\ref{eq:metri}).
The second line in the above equation contains all the inertial tidal
forces and Coriolis forces, the third line is the tidal force from the spacetime curvature,
where the tidal matrix is defined as
\begin{subequations}
\begin{eqnarray}
K_{\nu}^{\ \mu} & = & R_{\rho\nu\lambda}^{\ \ \ \ \mu}\tau^{\rho}\tau^{\lambda},\label{eq:Kdefinition}\\
K_{(a)(b)} & = & K_{\text{\ensuremath{\mu\nu}}}e_{(a)}^{\ \ \ \mu}e_{(b)}^{\ \ \ \nu}.\label{eq:Klocal}
\end{eqnarray}
\end{subequations}
For electrostatic and superconducting gradiometers, the motions of
test masses are suppressed by compensating forces. Then the total
tidal tensor $T_{(a)(b)}$ affecting the gradiometer will be
\begin{equation}
T_{(a)(b)}=-\frac{d}{d\tau}\gamma_{(a)(b)(0)}-\gamma_{(a)(c)(0)}\gamma_{\ \ (b)(0)}^{(c)}-K_{(a)(b)}.\label{eq:Tab}
\end{equation}

Along the orbit, Eqs. (\ref{eq:orbit}), we derive the
tidal matrix $K_{\nu}^{\ \mu}$ in the Earth-centered PN coordinate
system, which can be divided into the Newtonian $K^{N}$, the 1PN
GE $K^{GE}$, the GM $K^{GM}$ and the secular $K^{LT}$ parts:
\begin{widetext}
\begin{subequations}\label{eq:K}
\begin{equation}
(K^{N})_{\mu\nu}=\frac{M}{a^{3}}\left(\begin{array}{cccc}
0 & 0 & 0 & 0\\
\\
0 & -\frac{1}{2}(3\cos2\Psi+1) & -\frac{3}{2}\cos i\sin2\Psi & -\frac{3}{2}\sin
i\sin2\Psi\\
\\
0 & -\frac{3}{2}\cos i\sin2\Psi & \frac{1}{4}(6\cos2\Psi\cos^{2}i-3\cos2i+1) &
-3\cos i\sin i\sin^{2}\Psi\\
\\
0 & -\frac{3}{2}\sin i\sin2\Psi & -3\cos i\sin i\sin^{2}\Psi &
\frac{1}{4}(6\cos2\Psi\sin^{2}i+3\cos2i+1)
\end{array}\right),\label{eq:KN}
\end{equation}
\begin{eqnarray}
&&(K^{GE})_{\mu\nu}\nonumber\\
&&= \frac{M}{a^{3}}  \left(\begin{array}{cccc}
a^{2}\omega^{2} & a\omega\sin\Psi & -a\omega\cos i\cos\Psi & -a\omega\sin
i\cos\Psi\\
\\
a\omega\sin\Psi & \frac{M}{2a}(1+3\cos2\Psi) &
(\frac{3M}{2a}-2a^{2}\omega^{2})\cos i\sin2\Psi &
(\frac{3M}{2a}-2a^{2}\omega^{2})\sin i\sin2\Psi\\
 & -a^{2}\omega^{2}(1+2\cos2\Psi)\\
\\
-a\omega\cos i\cos\Psi & (\frac{3M}{2a}-2a^{2}\omega^{2})\cos i\sin2\Psi &
a^{2}\omega^{2}(2\cos2\Psi\cos^{2}i-2\cos2i+1) &
a^{2}\omega^{2}\sin2i(\cos2\Psi-2)\\
 &  & +\frac{M(-6\cos2\Psi\cos^{2}i+3\cos2i-1)}{4a} &
+\frac{3M\sin2i\sin^{2}\Psi}{2a}\\
\\
-a\omega\sin i\cos\Psi & (\frac{3M}{2a}-2a^{2}\omega^{2})\sin i\sin2\Psi &
a^{2}\omega^{2}\sin2i(\cos2\Psi-2) &
a^{2}\omega^{2}(2\cos2\Psi\sin^{2}i+2\cos2i+1)\\
 &  & +\frac{3M\sin2i\sin^{2}\Psi}{2a} &
-\frac{M(6\cos2\Psi\sin^{2}i+3\cos2i+1)}{4a}
\end{array}\right),\nonumber \\
\label{eq:KGE}
\end{eqnarray}
\begin{eqnarray}
&& (K^{GM})_{\mu\nu}\nonumber\\
&&=\frac{3J\omega}{a^{3}}\left(\begin{array}{cccc}
0 & 0 & 0 & 0\\
\\
0 & 2\cos i\cos^{2}\Psi & \frac{1}{2}(3\cos2i-1)\sin2\Psi &
3\sin2i\sin\Psi\cos\Psi\\
\\
0 & \frac{1}{2}(3\cos2i-1)\sin2\Psi & \cos i(5\cos2i-3)\sin^{2}\Psi &
\frac{1}{4}(10\sin3i\sin^{2}\Psi+3\cos2\Psi\sin i+\sin i)\\
\\
0 & 3\sin2i\sin\Psi\cos\Psi & \frac{1}{4}(10\sin3i\sin^{2}\Psi+3\cos2\Psi\sin
i+\sin i) & -\frac{1}{4}(5\cos3i+\cos i(20\cos2\Psi\sin^{2}i+3))
\end{array}\right),\nonumber \\
\label{eq:KGM}
\end{eqnarray}
\begin{eqnarray}
(K^{LT})_{\mu\nu}=\frac{3JM\Psi}{a^{6}\omega}\left(\begin{array}{cccc}
0 & 0 & 0 & 0\\
\\
0 & 2\cos i\sin2\Psi & -\frac{1}{2}(2\sin^{2}i+(\cos2i+3)\cos2\Psi) &
\sin2i\sin^{2}\Psi\\
\\
0 & -\frac{1}{2}(2\sin^{2}i+(\cos2i+3)\cos2\Psi) & -2\cos i\sin2\Psi & -\sin
i\sin2\Psi\\
\\
0 & \sin2i\sin^{2}\Psi & -\sin i\sin2\Psi & 0
\end{array}\right).\nonumber\\
\label{eq:KS}
\end{eqnarray}
\end{subequations}
\end{widetext}This secular tidal tensor Eq. (\ref{eq:KS}) is of the
1PN level: $(K^{LT})_{\nu}^{\ \mu}\sim\frac{JM\tau}{a^{6}}\sim\frac{1}{a^{2}}\Psi\mathcal{O}(\epsilon^{4})$
during the mission life time $\Psi\sim10^{4}$, which is produced
by the LT precession of the orbit relative to the Earth-centered PN coordinate
system.

\section{Tidal tensor under inertially guided local frame \label{sec:The-Fermi}}

Suppose the S/C is in the inertially guided orientation, which means that
the local tetrad $E_{(a)}^{\ \ \ \mu}$ attached to the S/C is non-rotating
and is parallel-shifted (Fermi-shifted) along the orbit. This defines
the local Fermi normal frame for the gradient measurements, in
which the geodesic deviation equation has a rather simple form:
\begin{equation}
\frac{d^{2}}{d\tau^{2}}Z^{(a)}=-K_{(b)}^{\ (a)}Z^{(b)}.\label{eq:dev_Fermi}
\end{equation}
From the tidal tensor in the Earth-centered PN system, Eqs. (\ref{eq:K}),
we have the order estimation :
\begin{equation}
K_{\nu}^{\ \mu}\sim\frac{1}{a^{2}}(\mathcal{O}(\epsilon^{2})+\mathcal{O}(\epsilon^{3})+\mathcal{O}(\epsilon^{4})+\Psi\mathcal{O}(\epsilon^{4}))\label{eq:Korder}.
\end{equation}
Therefore, to evaluate the tidal tensor in the local Fermi frame,
we only need to find the Fermi-shifted bases to the 1PN level:
\begin{equation}
E_{(a)}^{\ \ \ \mu}\sim\mathcal{O}(1)+\mathcal{O}(\epsilon^{2})+\Psi\mathcal{O}(\epsilon^{2}).\label{eq:Eorder}
\end{equation}

\subsection{Fermi-shifted tetrad}

First, in the Earth-centered PN coordinate system, we work out how
the spatial bases $\vec{E}_{(i)}$ are propagated freely along
the orbit, Eqs. (\ref{eq:orbit}), and then with the boost
transformations we obtain the Fermi-shifted tetrad attached to the
freely falling S/C. At the initial point $\vec{x}(0)=\{a,\ 0,\ 0\}$,
we set the origin of the Fermi normal frame to sit at the mass center
of the S/C, and the spatial basis $\vec{E}_{(1)}(0)$ along the initial
direction of motion of the reference mass, $\vec{E}_{(2)}(0)$ along
the initial radial direction, and $\vec{E}_{(3)}(0)=\vec{E}_{(1)}(0)\times\vec{E}_{(2)}(0)$
along the transverse direction:
\begin{subequations}\label{seq:E0}
\begin{eqnarray}
\vec{E}_{(1)}(0) & = & (1-\frac{M}{a})\left(\begin{array}{c}
0\\
\cos i\\
\sin i
\end{array}\right),\label{eq:E10}\\
\vec{E}_{(2)}(0) & = & (1-\frac{M}{a})\left(\begin{array}{c}
1\\
0\\
0
\end{array}\right),\label{eq:E20}\\
\vec{E}_{(3)}(0) & = & (1-\frac{M}{a})\left(\begin{array}{c}
0\\
\sin i\\
-\cos i
\end{array}\right).\label{eq:E30}
\end{eqnarray}
\end{subequations}
According to GR, such bases, which freely propagate along the orbit,
will undergo the relativistic precession with angular velocity \cite{Schiff1960}:
\begin{equation}
\vec{\Omega}=\frac{3M}{2a^{3}}\vec{x}\times\vec{v}-\frac{(a^{2}\vec{J}-3(\vec{J}\cdot\vec{x})\vec{x})}{a^{5}},\label{eq:Omega}
\end{equation}
where the first term corresponds to the geodetic precession and the
second term the frame-dragging precession. At the 1PN level and
in the limit $\frac{t}{a}\ll\frac{a^{2}}{J}$, the precession of the
normalized Fermi tetrad can be shown as
\begin{subequations}
\begin{eqnarray}
\vec{E}_{(1)}(\tau) & = & \left(\begin{array}{c}
\frac{J\tau\cos i}{a^{3}}-\sin\left(\frac{3M\Psi}{2a}\right)\\
\\
\left(1-\frac{M}{a}\right)\cos\left(\frac{3M\Psi}{2a}\right)\cos i-\frac{3J\sin^{2}i\sin^{2}\Psi}{2a^{3}\omega}\\
\\
\left(1-\frac{M}{a}\right)\cos\left(\frac{3M\Psi}{2a}\right)\sin i+\frac{3J\sin i\cos i\sin^{2}\Psi}{2a^{3}\omega}
\end{array}\right),\nonumber\\
\label{eq:vE1}\\
\vec{E}_{(2)}(\tau) & = & \left(\begin{array}{c}
\left(1-\frac{M}{a}\right)\cos\left(\frac{3M\Psi}{2a}\right)\\
\\
\sin\left(\frac{3M\Psi}{2a}\right)\cos i-\frac{J(3\sin^{2}i\sin2\Psi+3\Psi\cos2i+\Psi)}{4a^{3}\omega}\\
\\
\sin\left(\frac{3M\Psi}{2a}\right)\sin i+\frac{3J\sin2i(\sin2\Psi-2\Psi)}{8a^{3}\omega}
\end{array}\right),\nonumber\\
\label{eq:vE2}\\
\vec{E}_{(3)}(\tau) & = & \left(\begin{array}{c}
\frac{J\sin i(3\sin2\Psi-2\Psi)}{4a^{3}\omega}\\
\\
\left(1-\frac{M}{a}\right)\sin i+\frac{3J\sin i\cos i\sin^{2}\Psi}{2a^{3}\omega}\\
\\
\left(\frac{M}{a}-1\right)\cos i+\frac{3J\sin^{2}i\sin^{2}\Psi}{2a^{3}\omega}
\end{array}\right)\label{eq:vE3}.
\end{eqnarray}
\end{subequations}
One can see that the bases $\vec{E}_{(1)}$ and $\vec{E}_{(2)}$ are rotating
in the orbital plane with period $\frac{4\pi a}{3M\omega}$ due to
the geodetic effect, while the three bases are all following nutational
oscillations and precession around the $\vec{J}$ direction because
of the frame-dragging effect. This is \textit{not a new relativistic effect}
caused by a rotating source.

By definition, $E_{(0)}^{\ \ \ \mu}=\tau^{\mu}$, and
\begin{equation}
\tau^{\mu}=\left(\begin{array}{c}
\frac{dt}{d\tau}\\
\\
-a\omega\sin\Psi-\frac{2J\cos i(\sin\Psi+\Psi\cos\Psi)}{a^{2}}\\
\\
a\omega\cos i\cos\Psi+\frac{2J(\cos\Psi-\Psi\sin\Psi)}{a^{2}}\\
\\
a\omega\sin i\cos\Psi
\end{array}\right).\label{eq:Z}
\end{equation}
From the differential line
element $d\tau^{2}=-g_{\mu\nu}dx^{\mu}dx^{\nu}$ along the orbit,
we have
\begin{eqnarray}
\frac{dt}{d\tau} & = & 1+\frac{a^{2}\omega^{2}}{2}+\frac{M}{a}-\frac{a^{4}\omega^{4}}{8}+\frac{3Ma\omega^{2}}{2}+\frac{M^{2}}{2a^{2}}.\label{eq:t/tau}
\end{eqnarray}
Then, with the boost transformation, the spatial 1PN Fermi-shifted
tetrad attached to the S/C reads
\begin{subequations}\label{seq:Emu}
\begin{eqnarray}
E_{(1)}^{\ \ \ \mu} & = & \left(\begin{array}{c}
a\omega\cos\left(\Psi-\frac{3M\Psi}{2a}\right)\\
\\
-\frac{1}{2}a^{2}\omega^{2}\sin\Psi\cos\left(\Psi-\frac{3M\Psi}{2a}\right)\\
-\sin\left(\frac{3M\Psi}{2a}\right)+\frac{J\tau\cos i}{a^{3}}\\
\\
\frac{1}{2}a^{2}\omega^{2}\cos i\cos\Psi\cos\left(\Psi-\frac{3M\Psi}{2a}\right)\\
+(1-\frac{M}{a})\cos i\cos\left(\frac{3M\Psi}{2a}\right)\\
-\frac{3J\sin^{2}i\sin^{2}\Psi}{2a^{3}\omega}\\
\\
\frac{1}{4}a^{2}\omega^{2}\sin i\cos\left(2\Psi-\frac{3M\Psi}{2a}\right)\\
+(1-\frac{M}{a}+\frac{1}{2}a^{2}\omega^{2})\sin i\cos\left(\frac{3M\Psi}{2a}\right)\\
+\frac{3J\sin i\cos i\sin^{2}\Psi}{2a^{3}\omega}
\end{array}\right),\nonumber \\
\label{eq:E1}
\end{eqnarray}
\begin{eqnarray}
E_{(2)}^{\ \ \ \mu} & = & \left(\begin{array}{c}
-a\omega\sin\left(\Psi-\frac{3M\Psi}{2a}\right)\\
\\
\frac{1}{2}a^{2}\omega^{2}\sin\Psi\sin\left(\Psi-\frac{3M\Psi}{2a}\right)\\
+(1-\frac{M}{a})\cos\left(\frac{3M\Psi}{2a}\right)\\
\\
-\frac{1}{2}a^{2}\omega^{2}\cos i\cos\Psi\sin\left(\Psi-\frac{3M\Psi}{2a}\right)\\
+\cos i\sin\left(\frac{3M\Psi}{2a}\right)\\
-\frac{3J\sin^{2}i\sin2\Psi+J(3\Psi\cos2i+\Psi)}{4a^{3}\omega}\\
\\
-\frac{1}{4}a^{2}\omega^{2}\sin i\sin\left(2\Psi-\frac{3M\Psi}{2a}\right)\\
+\left(1+\frac{1}{4}a^{2}\omega^{2}\right)\sin i\sin\left(\frac{3M\Psi}{2a}\right)\\
-\frac{3J\sin i\cos i(\sin2\Psi-2\Psi)}{4a^{3}\omega}
\end{array}\right),\nonumber \\
\label{eq:E2}
\end{eqnarray}
\begin{equation}
E_{(3)}^{\ \ \ \mu}=\left(\begin{array}{c}
\frac{J\sin i(4\cos\Psi-3\Psi\sin\Psi)}{2a^{2}}\\
\\
\frac{J\sin i(3\sin2\Psi-2\Psi)}{4a^{3}\omega}\\
\\
\sin i\left(1-\frac{M}{a}+\frac{3J\cos i\sin^{2}\Psi}{2a^{3}\omega}\right)\\
\\
\left(\frac{M}{a}-1\right)\cos i+\frac{3J\sin^{2}i\sin^{2}\Psi}{2a^{3}\omega}
\end{array}\right).\label{eq:E3}
\end{equation}
\end{subequations}

\subsection{Tidal tensor in inertially guided system}

With Eqs. (\ref{eq:K}) and the explicit expressions
of the Fermi-shifted tetrad, Eqs. (\ref{eq:Z})-(\ref{seq:Emu}), we can
directly work out the total tidal tensor $K_{(a)(b)}$ in the inertially
guided local frame along the orbit. The components $K_{(0)(a)}=0$,
as expected at 1PN level. Therefore, the spatial parts of the Newtonian,
1PN GE and GM tidal tensors read
\begin{subequations}\label{eq:KabPN}
\begin{equation}
K_{(i)(j)}^{N}=\frac{M}{a^{3}}\left(\begin{array}{ccc}
\frac{3\cos2\Psi-1}{2} & -\frac{3\sin2\Psi}{2} & 0\\
\\
-\frac{3\sin2\Psi}{2} & -\frac{3\cos2\Psi+1}{2} & 0\\
\\
0 & 0 & 1
\end{array}\right),\label{eq:KabN}
\end{equation}
\begin{eqnarray}
 &  & K_{(i)(j)}^{GE}=\nonumber \\
 &  & \frac{3M}{a^{3}}\left(\begin{array}{ccc}
\frac{M(1-3\cos2\Psi)}{2a} & \frac{3M\sin2\Psi}{2a}& 0\\
-a^{2}\omega^{2}\sin^{2}\Psi & -\frac{a^{2}\omega^{2}\sin2\Psi}{2} \\
\\
\frac{3M\sin2\Psi}{2a} & \frac{M(1+3\cos2\Psi)}{2a} & 0\\
 -\frac{a^{2}\omega^{2}\sin2\Psi}{2}& -a^{2}\omega^{2}\cos^{2}\Psi\\
\\
0 & 0 & a^{2}\omega^{2}-\frac{M}{a}
\end{array}\right),\nonumber \\
\label{eq:KabGE}
\end{eqnarray}
{\small
\begin{eqnarray}
 &  & K_{(i)(j)}^{GM}=\nonumber \\
 &  & \frac{3J\omega}{a^{3}}\left(\begin{array}{ccc}
2\cos i\sin^{2}\Psi & \cos i\sin2\Psi & \sin i(2\cos2\Psi-1)\\
\\
\cos i\sin2\Psi & 2\cos i\cos^{2}\Psi & -2\sin i\sin2\Psi\\
\\
\sin i(2\cos2\Psi-1) & -2\sin i\sin2\Psi & -2\cos i
\end{array}\right).\nonumber \\
\label{eq:KabGM}
\end{eqnarray}}
\end{subequations}These results \textit{agree exactly} with the tidal tensors evaluated \textit{in
the non-precessing frame} along Keplerian orbits by Mashhoon, Paik and Will \cite{Mashhoon1989}.

\begin{figure}
\includegraphics[scale=0.38]{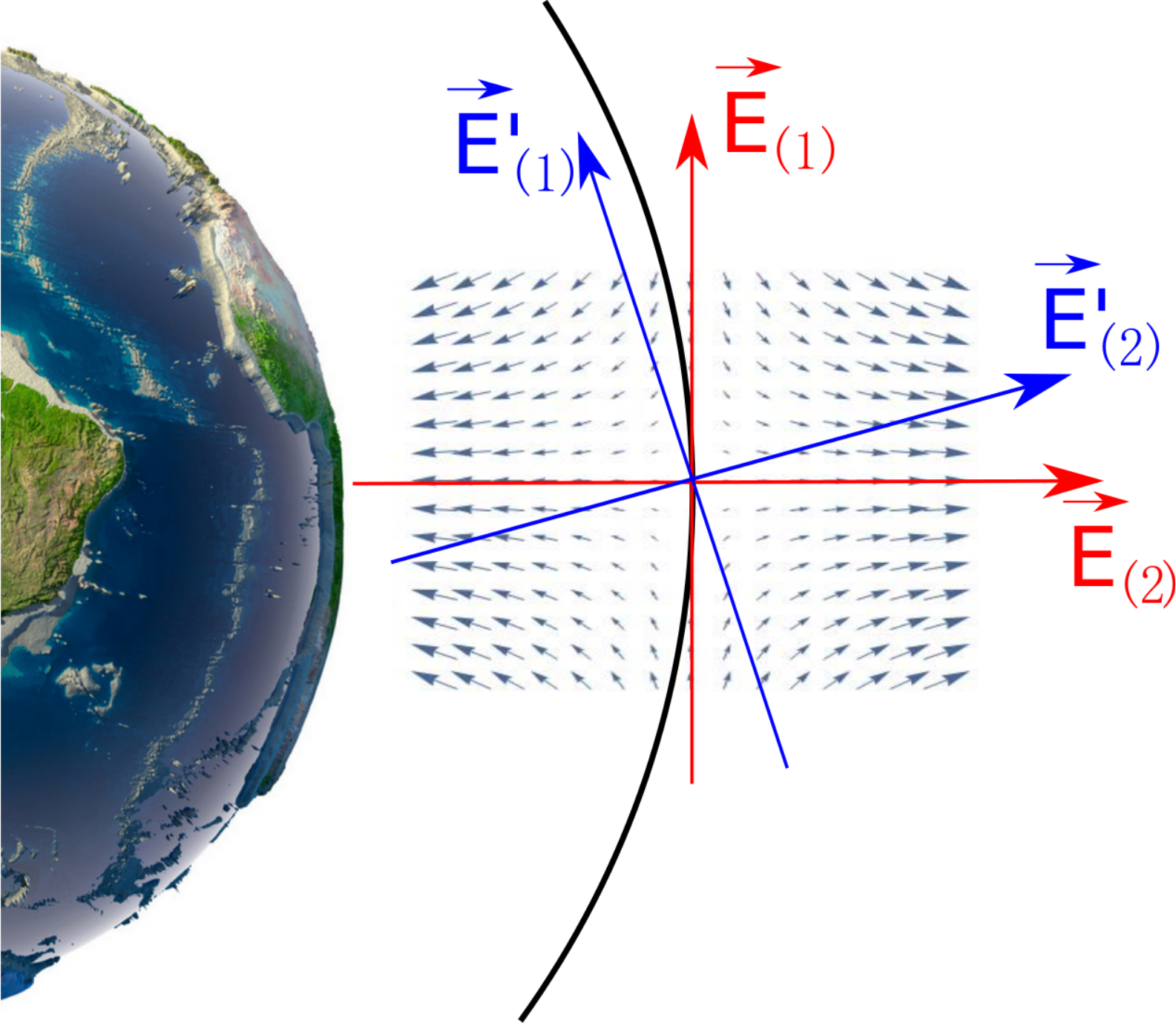}
\caption{Take the ascending node as the sample point. In the orbital plane,
the Newtonian tidal field at this point are plotted. Due to the geodetic
effect, the bases $\vec{E}_{(1)}$ and $\vec{E}_{(2)}$ will precess
to $\vec{E}'_{(1)}$ and $\vec{E}'_{(2)}$ after one turn along the
orbit. Therefore, the Newtonian tidal forces along the bases will
change gradually with time due to the continuous precession of the
orientation of the bases relative to the local Newtonian tidal field,
which, in the limit $\frac{\tau}{a}\ll\frac{1}{M\omega}$,
will result in the geodetic part of the secular tidal forces.}
\label{fig:geodetic}
\end{figure}

\begin{figure}
\includegraphics[scale=0.38]{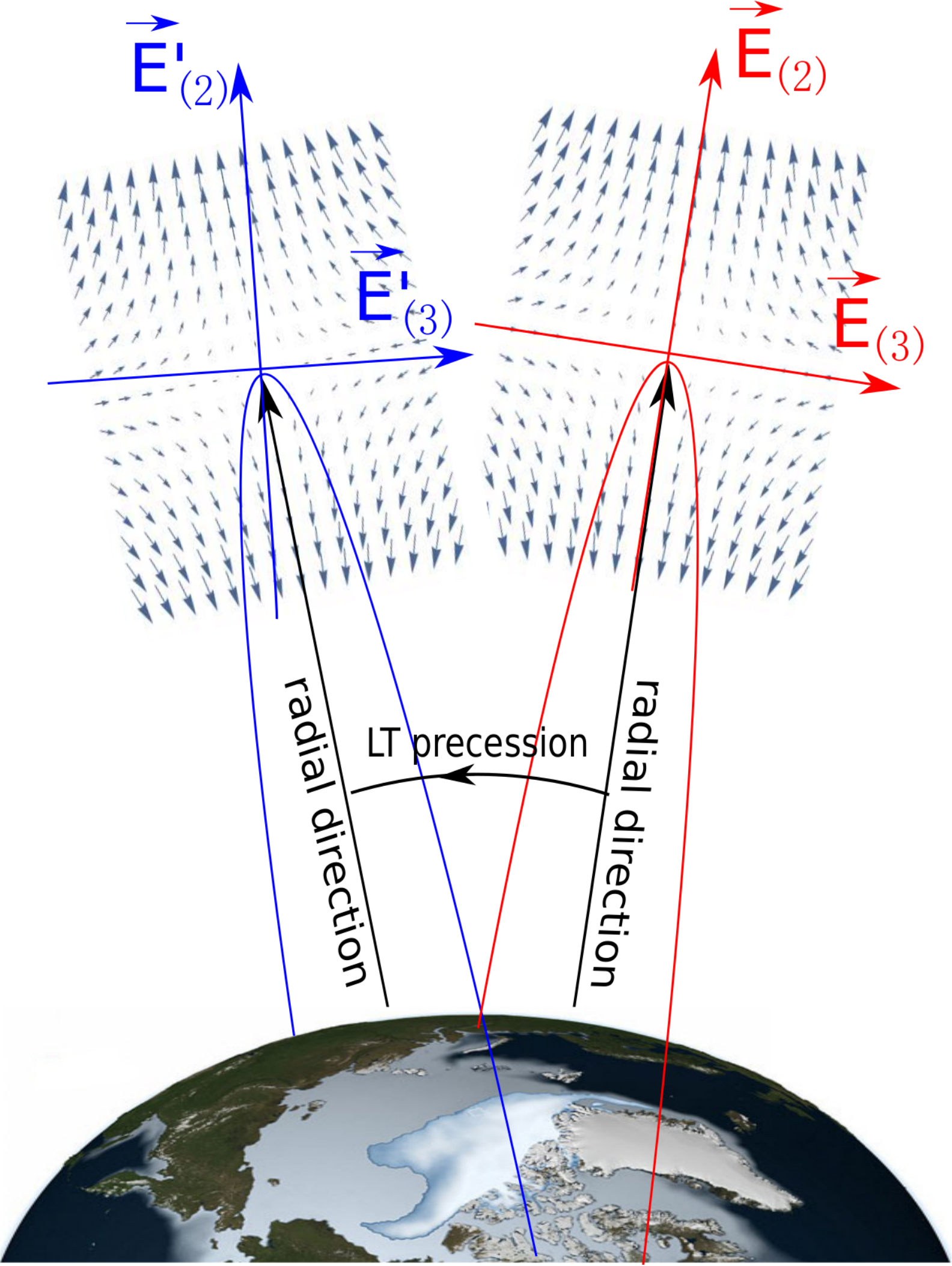}
\caption{Again, take the ascending nodes as sample points, whose location will
precess along the equator due to the LT precession of the orbit. In
the equatorial plane, the Newtonian tidal fields at these two ascending
nodes are plotted. Due to the frame-dragging effect, the bases will
undergo the Schiff precession with an angular rate \textit{different} from
that of the LT precession of the orbit. Therefore, the orientation
of the bases relative to the local Newtonian tidal field will be altered
gradually, which, in the limit $\frac{\tau}{a}\ll\frac{a^{2}}{J}$,
produces secular changes of the Newtonian tidal forces along these
bases.}
\label{fig:framedragging}
\end{figure}

In this paper, we extend the analysis by \textit{including the frame-dragging effect}.
Unlike in \cite{Mashhoon1989} which considered the inertial frame defined by remote stars,
we now consider the \textit{local inertial frame} defined by a set of on-board gyros.
Due to the precessing rate difference
between the local frame and the orbital plane, the relative orientation
between the local freely falling frame and the sidereal frame will change gradually with time. Therefore, the
Newtonian tidal forces along the bases $E_{(i)}^{\ \ \ \mu}(\tau)$
will be modulated by such an orientation change, which results in
the secular parts of the total tidal tensor. Such secular tidal effects
can be divided into two different parts according to their origins:
the secular tensors $K_{(a)(b)}^{G}$ and $K_{(a)(b)}^{FD}$
generated respectively by the geodetic precession and frame-dragging
precession. This is illustrated in Figs. \ref{fig:geodetic} and \ref{fig:framedragging}.

At the 1PN level, and in the limit compared to the
periods of the geodetic and frame-dragging precession, that $\frac{\tau}{a}\ll\frac{1}{M\omega}$
and $\frac{\tau}{a}\ll\frac{a^{2}}{J}$, we have
\begin{subequations}\label{eq:KSecular}
\begin{equation}
K_{(i)(j)}^{G}=\frac{9M^{2}\Psi}{2a^{4}}\left(\begin{array}{ccc}
\sin2\Psi & \cos2\Psi & 0\\
\\
\cos2\Psi & -\sin2\Psi & 0\\
\\
0 & 0 & 0
\end{array}\right),\label{eq:Kgeodetic}
\end{equation}
\begin{widetext}
\begin{eqnarray}
&&K_{(i)(j)}^{FD} \nonumber\\
& &=  \frac{9JM}{a^{6}\omega}\left(\begin{array}{ccc}
-\Psi\cos i\sin2\Psi & -\Psi\cos i\cos2\Psi & -\frac{1}{4}\sin i(\Psi\sin2\Psi-\cos2\Psi+1)\\
\\
-\Psi\cos i\cos2\Psi & \Psi\cos i\sin2\Psi & -\frac{1}{4}\sin i[\Psi 
(\cos2\Psi+1)+\sin2\Psi] \\
\\
-\frac{1}{4}\sin i(\Psi\sin2\Psi-\cos2\Psi+1) & -\frac{1}{4}\sin i[\Psi 
(\cos2\Psi+1)+\sin2\Psi] & 0
\end{array}\right).\nonumber \\
\label{eq:KFD}
\end{eqnarray}
\end{widetext}
 \end{subequations}The above secular tidal tensor $K^{FD}_{(i)(j)}$ from
frame-dragging
effect \textit{does not agree exactly}
with the results from Mashhoon and Theiss
\cite{Mashhoon1982,Mashhoon1985,Theiss1985}
that are summarized in Eq. (12) of \cite{Paik2008}. To check the
validity of these results,
one should notice that, when the local, freely falling frame and the
sidereal frame coincide at the initial point $\vec{x}=0$ and $\Psi=0$,
the secular tensors $K^G_{(i)(j)}$ and $K^{FD}_{(i)(j)}$ should
vanish. This condition is exactly satisfied by our result when
$\Psi=0$ is substituted into Eqs. (\ref{eq:Kgeodetic}) and
(\ref{eq:KFD}). However, such a self-consistency check is not fully
satisfied for the results from Mashhoon and Theiss.

\section{Tidal tensor in the Earth-pointing local frame\label{sec:Earth-pointing}}

The Earth-pointing orientation of the S/C (local frame) is another feasible
option for satellite gradiometry.
Strictly speaking, at the 1PN level, the exact freely falling
Earth-pointing frame exists only for polar orbits. Along inclined orbits,
there will always be periodic nutations of the local frame (S/C) within
the orbital plane due to the frame-dragging effect. Therefore, without
attitude control, freely falling Earth-pointing frame is only an approximation
at the 1PN level.

Now, we first work out the 1PN tetrad $e_{(a)}^{\ \ \ \mu}$ determining
the approximate Earth-pointing frame along the orbit, Eqs .(\ref{eq:orbit}).
The initial values of the spatial bases $\vec{e}_{(i)}(0)$ are set
to be the same as in Eqs. (\ref{seq:E0}). To obtain
its Earth-pointing orientation approximately, the local frame (S/C) is set to roll
about the transverse axis $\vec{e}_{(3)}$ with an initial angular
velocity $w_{0}$ relative to the Fermi normal frame. Considering
the uniform geodetic precession of the local frame, we set
\begin{equation}
w_{0}=\omega(1-\frac{3M}{2a}).\label{eq:w0}
\end{equation}
Then, following the same method used in the last section, the matrix $e_{(a)}^{\ \ \ \mu}$
and its inverse $e_{\ \ \ \mu}^{(a)}$ can be derived to the required
order, the result is given in Eqs. (\ref{eq:e}) and
(\ref{eq:ie}) in Appendix. \ref{sec:Christoffel-symbols-and}

From Eq. (\ref{eq:Tab}), $T_{(0)(a)}=0$, and the Newtonian, 1PN GE
and GM parts of the total tidal tensor $T_{(i)(j)}$ are found to be
\begin{subequations}\label{eq:TPN}
\begin{equation}
T^{N}{}_{(i)(j)}=\left(\begin{array}{ccc}
\frac{M}{a^{3}}-\omega^{2} & 0 & 0\\
\\
0 & -\omega^{2}-\frac{2M}{a^{3}} & 0\\
\\
0 & 0 & \frac{M}{a^{3}}
\end{array}\right),\label{eq:TN}
\end{equation}
\begin{eqnarray}
 &  & T^{GE}{}_{(i)(j)}=\nonumber \\
 &  & \left(\begin{array}{ccc}
-\frac{\omega^{4}a^{6}-4M\omega^{2}a^{3}+3M^{2}}{a^{4}} & 0 & 0\\
\\
0 & \frac{7M^2-M\omega^{2}a^3}{a^{4}} & 0\\
\\
0 & 0 & \frac{3M\omega^{2}a^3-3M^2}{a^{4}}
\end{array}\right),\nonumber \\
\label{eq:TGE}
\end{eqnarray}
\begin{eqnarray}
 &  & T^{GM}{}_{(i)(j)}=\nonumber \\
 &  & \frac{3J\omega}{a^{3}}\left(\begin{array}{ccc}
0 & 0 & \sin i\cos\Psi\\
\\
0 & 2\cos i & -3\sin i\sin\Psi\\
\\
\sin i\cos\Psi & -3\sin i\sin\Psi & -2\cos i
\end{array}\right).\nonumber \\
\label{eq:TGM}
\end{eqnarray}
\end{subequations}
Again, these results agree \textit{to the 1PN level} with the tidal tensors obtained by Mashhoon, Paik and Will \cite{Mashhoon1989} for the Earth-pointing frame. As the geodetic precession of the bases is absorbed into their rotations
about the axis $e_{(3)}^{\ \ \ \mu}$, there is no secular
part in $T_{(i)(j)}$ produced by the geodetic precession as for the
inertially guided case. However, an error $\delta\omega_{0}$ in the initial
rolling velocity of the S/C will produce a secular tidal field
of order $\mathcal{O}(\frac{M}{a^{3}}\tau\delta\omega_{0})$ in the $e_{(1)}^{\ \ \ \mu}-e_{(2)}^{\ \ \ \mu}$
plane (orbital plane), which could be used to adjust the attitude
of the S/C in that plane. The effects and errors produced by the uncertainties
in the angular velocity of the S/C relative to its mass center will
be left to future studies.

If the S/C is forced to strictly follow the Earth-pointing frame by attitude control,
there will be no secular terms.  However, if the attitude is controlled by on-board gyros,
there will be frame-dragging terms due to the Schiff precession of the gyro axes.
In the limit $\frac{\tau}{a}\ll\frac{a^{2}}{J}$,
the frame-dragging part of the secular tidal tensor can be shown to be
\begin{widetext}
\begin{eqnarray}
 T^{FD}{}_{(i)(j)}=
\frac{9JM}{a^{6}\omega}\left(\begin{array}{ccc}
0 & -\Psi\cos i & 0\\
\\
-\Psi\cos i & 0 & -\frac{1}{2}\sin i(\Psi\cos\Psi+\sin\Psi)\\
\\
0 & -\frac{1}{2}\sin i(\Psi\cos\Psi+\sin\Psi) & 0
\end{array}\right).\label{eq:TSFD}
\end{eqnarray}
\end{widetext}

\section{Concluding remarks\label{sec:Conclusions}}

In this paper, we have derived the complete 1PN tidal
tensor, including secular tensors, along 1PN nearly circular orbits under both  inertially
guided and Earth-pointing frames. At the 1PN level, these secular tidal forces
are not new (resonance) effects produced by a rotating source. From our
analysis, we find a simple physical picture: \textit{the Newtonian tidal forces
along certain bases are modulated
due to the relativistic precession of the local frame relative to
the sidereal frame,} which results in different secular terms
in the tidal tensor.
Nevertheless, the existence of the secular tidal terms does provide
the possibility of measuring the GM effect with a great accuracy in
satellite gradiometry missions \cite{Theiss1985}, although such an experiment would require rather precise gyros \cite{Mashhoon1982}.

\begin{figure}
\center\includegraphics[scale=0.5]{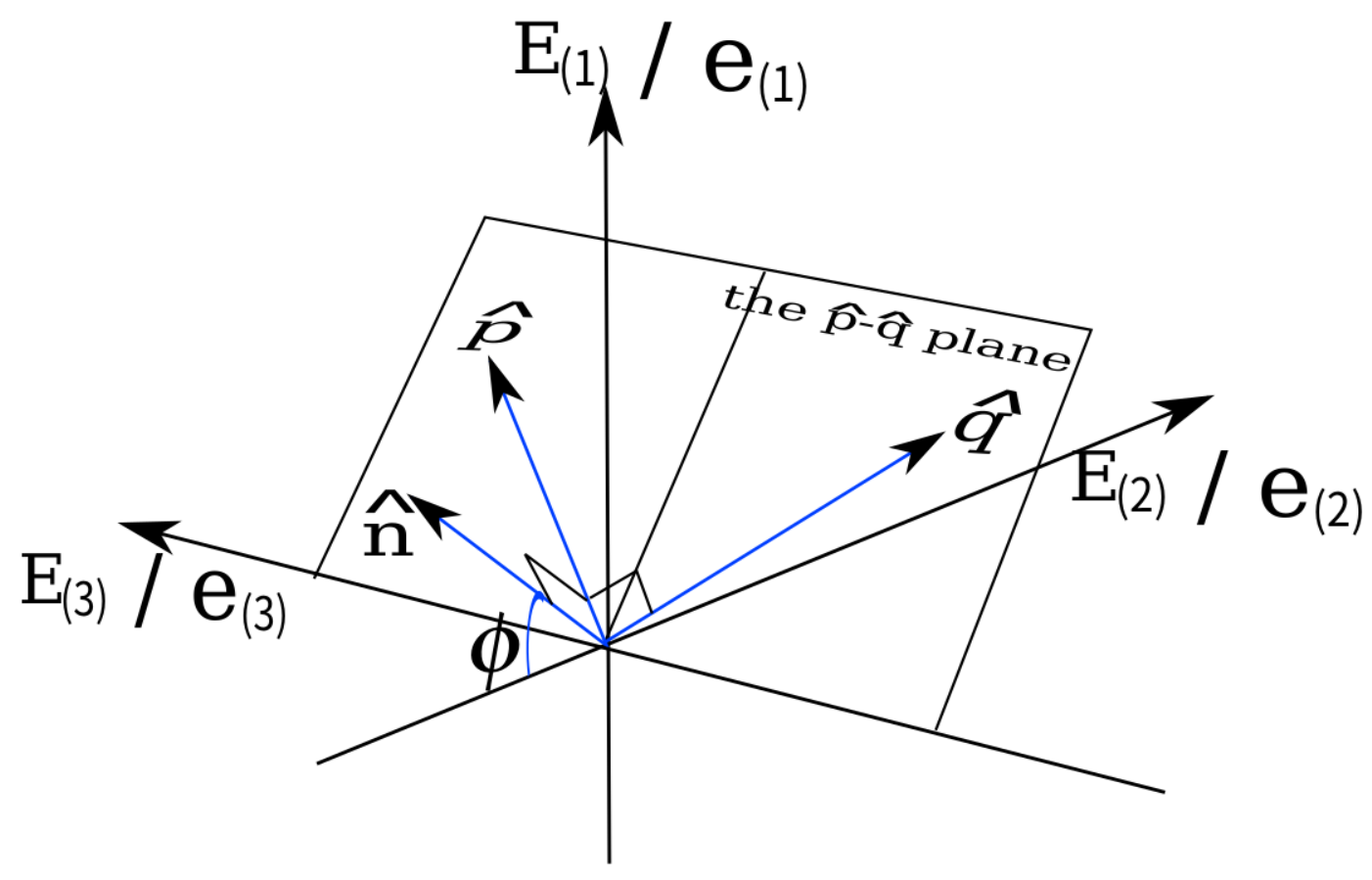}
\caption{The two perpendicular axes $\hat{\mathbf{p}}$ and $\hat{\mathbf{q}}$
are symmetric with respect to the $E_{(1)}^{\ \ \ \mu}-E_{(2)}^{\ \ \ \mu}$ or
$e_{(1)}^{\ \ \ \mu}-e_{(2)}^{\ \ \ \mu}$ plane, and $\hat{\mathbf{n}}$
is orthogonal to the $\hat{\mathbf{p}}-\hat{\mathbf{q}}$ plane. The
angle between $\hat{\mathbf{n}}$ and $-\hat{\mathbf{r}}$ is denoted
by $\phi$.}
\label{fig:3SGG}
\end{figure}
In conclusion, we
summarize here the gradiometer readout of the frame-dragging part of the secular tidal
tensor by a 3-axis gradiometer. Following \cite{Mashhoon1989}, we
orient two of the three gradiometer axes 45 degrees above and below the orbital
plane and difference their outputs to reject the Newtonian and GE terms and measure
only the GM and frame-dragging terms, as shown in Fig. \ref{fig:3SGG}. In the local frame $E_{(i)}^{\ \ \ \mu}$
or $e_{(i)}^{\ \ \ \mu}$, the three axes of the gradiometer, $\hat{\mathbf{n}},$
$\hat{\mathbf{p}}$ and $\hat{\mathbf{q}}$, are oriented as
\begin{subequations}
\begin{eqnarray}
\hat{\mathbf{n}} & = & \left(\begin{array}{c}
\sin\phi\\
-\cos\phi\\
0
\end{array}\right),\label{eq:n3}\\
\hat{\mathbf{p}} & = & \frac{1}{\sqrt{2}}\left(\begin{array}{c}
\cos\phi\\
\sin\phi\\
-1
\end{array}\right),\label{eq:p3}\\
\hat{\mathbf{q}} & = & \frac{1}{\sqrt{2}}\left(\begin{array}{c}
\cos\phi\\
\sin\phi\\
1
\end{array}\right).\label{eq:q3}
\end{eqnarray}
\end{subequations}
In the inertially guided and Earth-pointing frames, the frame-dragging
secular terms can be obtained in the difference between the readouts
in the $\hat{\mathbf{p}}$ and $\hat{\mathbf{q}}$ axes. For the inertially
guided case, the gradiometer measures
\begin{eqnarray}
 &  & \frac{1}{2}(K_{\hat{\mathbf{p}}\hat{\mathbf{p}}}-K_{\hat{\mathbf{q}}\hat{\mathbf{q}}})\nonumber \\
 & = & \boxed{\frac{9JM\Psi\sin i\cos\Psi\sin(\Psi+\phi)}{2a^{6}\omega}}\nonumber \\
 &  & -\frac{9JM\sin i\cos(2\Psi+\phi)}{4a^{6}\omega}-\frac{6J\omega\sin i\cos(2\Psi+\phi)}{a^{3}}\nonumber \\
 &  & +\frac{9JM\sin i\cos\phi}{4a^{6}\omega}+\frac{3J\omega\sin i\cos(\phi)}{a^{3}}.\label{eq:Kpq}
\end{eqnarray}
For the Earth-pointing case, the gradiometer measures
\begin{eqnarray}
 &  & \frac{1}{2}(T_{\hat{\mathbf{p}}\hat{\mathbf{p}}}-T_{\hat{\mathbf{q}}\hat{\mathbf{q}}})\nonumber \\
 & = & \boxed{\frac{9JM\Psi\sin i\sin\phi\cos\Psi}{2a^{6}\omega}}+\frac{9JM\sin i\sin\phi\sin\Psi}{2a^{6}\omega}\nonumber \\
 &  & +\frac{9J\omega\sin i\sin\phi\sin\Psi}{a^{3}}-\frac{3J\omega\sin i\cos\phi\cos\Psi}{a^{3}}.\label{eq:Tpq}
\end{eqnarray}
The boxed terms are the expected signals from secular terms. A detailed analysis
of the errors arising from misalignment and mispointing of the gradiometer axes can be
found in \cite{Paik2008}.

\begin{acknowledgments}
This work was supported in part by the National Natural Science Foundation of China under grants 11305255,
11171329 and 41404019 and by the National Science Foundation of USA
under grant PHY1105030. We are grateful to Yun Kau Lau for initiating the study of the problem and encouraging us to do this work.
We are also grateful to
Shing-Tung Yau and Le Yang for their continuous support
of our work at the Morningside Center of Mathematics, Chinese Academy of Sciences.
\end{acknowledgments}

\appendix
\section{Christoffel symbols and tetrad matrices\label{sec:Christoffel-symbols-and}}

As a reference, under the PN coordinate system defined in Sec. \ref{sec:Settings},
we give the expressions of the components of the Christoffel symbols
$\Gamma_{\:\:\nu\mu}^{\lambda}$:
\begin{subequations}\label{Gamma}
\begin{equation}
\Gamma_{\ \ \ 0\mu}^{0}=\frac{M}{r^{3}}\left(\begin{array}{c}
0\\
x^{1}\\
x^{2}\\
x^{3}
\end{array}\right),\label{Gamma00mu}
\end{equation}
\begin{eqnarray}
 &  & \Gamma_{\ \ \ 0j}^{i}=\nonumber\\
 &  & \left(\begin{array}{ccc}
0 & -\frac{J\left((x^{1})^{2}+(x^{2})^{2}-2(x^{3})^{2}\right)}{r^{5}} & -\frac{3Jx^{2}x^{3}}{r^{5}}\\
\\
\frac{J\left((x^{1})^{2}+(x^{2})^{2}-2(x^{3})^{2}\right)}{r^{5}} & 0 & \frac{3Jx^{1}x^{3}}{r^{5}}\\
\\
\frac{3Jx^{2}x^{3}}{r^{5}} & -\frac{3Jx^{1}x^{3}}{r^{5}} & 0
\end{array}\right),\nonumber \\
&&\label{Gammai0j}
\end{eqnarray}
\begin{equation}
\Gamma_{\ \ \ ij}^{1}=-\frac{M}{r^{3}}\left(\begin{array}{ccc}
x^{1} & x^{2} & x^{3}\\
x^{2} & -x^{1} & 0\\
x^{3} & 0 & -x^{1}
\end{array}\right),\label{Gamma1ij}
\end{equation}
\begin{equation}
\Gamma_{\ \ \ ij}^{2}=-\frac{M}{r^{3}}\left(\begin{array}{ccc}
-x^{2} & x^{1} & 0\\
x^{1} & x^{2} & x^{3}\\
0 & x^{3} & -x^{2}
\end{array}\right),\label{Gamma2ij}
\end{equation}
\begin{equation}
\Gamma_{\ \ \ ij}^{3}=-\frac{M}{r^{3}}\left(\begin{array}{ccc}
-x^{3} & 0 & x^{1}\\
0 & -x^{3} & x^{2}\\
x^{1} & x^{2} & x^{3}
\end{array}\right),\label{Gamma3ij}
\end{equation}
\begin{widetext}
\begin{eqnarray}
 &  & \Gamma_{\ \ \ ij}^{0}=\nonumber\\
 &  & \left(\begin{array}{ccc}
\frac{6J\left(x^{2}(x^{1})^{3}+x^{2}\left((x^{2})^{2}+(x^{3})^{2}\right)x^{1}\right)}{r^{7}} & \frac{3J\left(-(x^{3})^{2}(x^{1})^{2}-(x^{1})^{4}+(x^{2})^{2}\left((x^{2})^{2}+(x^{3})^{2}\right)\right)}{r^{7}} & \frac{3J\left(x^{2}x^{3}(x^{1})^{2}+x^{2}x^{3}\left((x^{2})^{2}+(x^{3})^{2}\right)\right)}{r^{7}}\\
\\
\\
\frac{3J\left(-(x^{3})^{2}(x^{1})^{2}-(x^{1})^{4}+(x^{2})^{2}\left((x^{2})^{2}+(x^{3})^{2}\right)\right)}{r^{7}} & -\frac{6J\left(x^{2}(x^{1})^{3}+x^{2}\left((x^{2})^{2}+(x^{3})^{2}\right)x^{1}\right)}{r^{7}} & -\frac{3J\left(x^{3}(x^{1})^{3}+x^{3}\left((x^{2})^{2}+(x^{3})^{2}\right)x^{1}\right)}{r^{7}}\\
\\
\\
\frac{3J\left(x^{2}x^{3}(x^{1})^{2}+x^{2}x^{3}\left((x^{2})^{2}+(x^{3})^{2}\right)\right)}{r^{7}} & -\frac{3J\left(x^{3}(x^{1})^{3}+x^{3}\left((x^{2})^{2}+(x^{3})^{2}\right)x^{1}\right)}{r^{7}} & 0
\end{array}\right).\nonumber\\
&&\label{Gamma0ij}
\end{eqnarray}
\end{widetext}
\end{subequations}

To the required order, the transformation matrices between the local
Earth-pointing frame and the Earth-centered PN coordinate system
$e_{(a)}^{\ \ \ \mu}$ and
$e_{\ \ \ \mu}^{(a)}$ can be shown to be
\begin{widetext}
\begin{eqnarray}
 &  & e_{(a)}^{\ \ \ \mu}=\nonumber \\
 &  & \left(\begin{array}{cccc}
1+\frac{a^{2}\omega^{2}}{2}+\frac{M}{a} & -a\omega\sin\Psi & a\omega\cos i\cos\Psi & a\omega\sin i\cos\Psi\\
 & -\frac{2J\cos i(\Psi\cos\Psi+\sin\Psi)}{a^{2}} & +\frac{2J(\cos\Psi-\Psi\sin\Psi)}{a^{2}}\\
\\
\\
(a+2M)\omega & -(1+\frac{a^{2}\omega^{2}}{2}-\frac{M}{a})\sin\Psi & (1+\frac{a^{2}\omega^{2}}{2}-\frac{M}{a})\cos i\cos\Psi & (1+\frac{a^{2}\omega^{2}}{2}-\frac{M}{a})\sin i\cos\Psi\\
 & +\frac{J\Psi\cos i\cos\Psi}{a^{3}\omega} & +\frac{J\Psi\left(1+3\cos i\right)\sin\Psi}{4a^{3}\omega} & +\frac{3J\Psi\sin i\sin\Psi}{4a^{3}\omega}\\
\\
\\
-\frac{3J\Psi\cos i}{a^{2}} & (1-\frac{M}{a})\cos\Psi & (1-\frac{M}{a})\cos i\sin\Psi & (1-\frac{M}{a})\sin i\sin\Psi\\
 & +\frac{J\Psi\cos i\sin\Psi}{a^{3}\omega} & -\frac{\left(1+3\cos2i\right)J\Psi\cos\Psi}{4a^{3}\omega}-\frac{3J\sin^{2}i\sin\Psi}{2a^{3}\omega} & -\frac{3J\sin2i\left(\Psi\cos\Psi-\sin\Psi\right)}{4a^{3}\omega}\\
\\
\\
-\frac{3J\Psi\sin i\sin\Psi}{2a^{2}} & \frac{J\sin i\left(3\sin2\Psi-2\Psi\right)}{4a^{3}\omega} & (1-\frac{M}{a})\sin i & -(1-\frac{M}{a})\cos i\\
 &  & +\frac{3J\sin i\cos i\sin^{2}\Psi}{2a^{3}\omega} & +\frac{3J\sin^{2}i\sin^{2}\Psi}{2a^{3}\omega}
\end{array}\right),\nonumber \\
\label{eq:e}
\end{eqnarray}
\begin{eqnarray}
 &  & e_{\ \ \ \mu}^{(a)}=\nonumber \\
 &  & \left(\begin{array}{cccc}
1+\frac{a^{2}\omega^{2}}{2}-\frac{M}{a}+\frac{2J\omega\cos i}{a} & -a\omega+\frac{M^{2}\omega}{a}-\frac{2J\cos i}{a^{2}} & \frac{3J\Psi\cos i}{a^{2}} & \frac{J\sin i(3\Psi\sin\Psi-4\cos\Psi)}{2a^{2}}\\
\\
\\
(a+2M)\omega\sin\Psi & -(1+\frac{a^{2}\omega^{2}}{2}+\frac{M}{a})\sin\Psi & (1+\frac{M}{a})\cos\Psi & \frac{3J\sin i\sin\Psi}{4a^{3}\omega}-\frac{J\Psi\sin i}{2a^{3}\omega}\\
+\frac{2J\Psi\cos i\cos\Psi}{a^{2}} & +\frac{J\Psi\cos i\cos\Psi}{a^{3}\omega} & +\frac{J\Psi\cos i\sin\Psi}{a^{3}\omega}\\
\\
\\
-(a+2M)\omega\cos i\cos\Psi & (1+\frac{a^{2}\omega^{2}}{2}+\frac{M}{a})\cos i\cos\Psi & (1+\frac{M}{a})\cos i\sin\Psi & (1+\frac{M}{a})\sin i\\
+\frac{2J\Psi\sin\Psi}{a^{2}} & +\frac{3J\Psi\cos2i\sin\Psi}{4a^{3}\omega}+\frac{J\Psi\sin\Psi}{4a^{3}\omega} & -\frac{3J\sin^{2}i\sin\Psi}{2a^{3}\omega}-\frac{J\Psi(3\cos2i+1)\cos\Psi}{4a^{3}\omega} & +\frac{3J\sin i\cos i\sin^{2}\Psi}{2a^{3}\omega}\\
\\
\\
-(a+2M)\omega\sin i\cos\Psi & (1+\frac{a^{2}\omega^{2}}{2}+\frac{M}{a})\sin i\cos\Psi & (1+\frac{M}{a})\sin i\sin\Psi & -(1+\frac{M}{a})\cos i\\
 & +\frac{3J\Psi\sin i\cos i\sin\Psi}{2a^{3}\omega} & -\frac{3J\sin i\cos i(3\Psi\cos\Psi-\sin\Psi)}{2a^{3}\omega} & +\frac{3J\sin^{2}i\sin^{2}\Psi}{2a^{3}\omega}
\end{array}\right).\nonumber \\
\label{eq:ie}
\end{eqnarray}
\end{widetext}


\end{document}